\documentclass[%
 reprint,
nofootinbib,
 amsmath,amssymb,
 aps,
pra,
]{revtex4-1}

\usepackage{graphicx}
\usepackage{dcolumn}
\usepackage{bm}
\usepackage[utf8]{inputenc}
\usepackage[T1]{fontenc}
\usepackage{mathptmx}
\usepackage{xcolor}
\usepackage{tikz}
\usetikzlibrary{fadings,shapes.arrows,shadows}  
\usetikzlibrary{arrows}
\usepackage{sansmath}
\usetikzlibrary{shadings,intersections}
\usepackage{verbatim}
\usepackage{subcaption}
\usepackage{caption}
\usepackage{hyperref}

\begin{document}

\preprint{APS/123-QED}

\title[]{Optomechanical transport of cold atoms induced by structured light}

\author{Giuseppe Baio\textsuperscript{1}}
\author{Gordon R. M. Robb\textsuperscript{1}}%
\author{Alison M. Yao\textsuperscript{1}}
\author{Gian-Luca Oppo\textsuperscript{1}\vspace*{0.2cm}}

\email{giuseppe.baio@strath.ac.uk}
\affiliation{
\textsuperscript{1}\textit{SUPA and Department of Physics, University of Strathclyde, Glasgow G4 0NG, Scotland, U.K.}}

\date{\today}

\begin{abstract}

Optomechanical pattern forming instabilities in a cloud of cold atoms lead to self-organized spatial structures of light and atoms.  Here, we consider the optomechanical self-structuring of a  cold atomic cloud in the presence of a phase structured input field, carrying orbital angular momentum. For a planar ring cavity setup, a model of coupled cavity field and atomic density equations describes a wide range of drifting modulation instabilities in the transverse plane. This leads to the formation of rotating self-organized rings of light-atom lattices. Using linear stability analysis and numerical simulations of the coupled atomic and optical dynamics, we demonstrate the presence of macroscopic atomic transport corresponding to the pattern rotation, induced by the structured pump phase profile.


\end{abstract}

\pacs{Valid PACS appear here}
\maketitle


\section{\label{sec:level1}Introduction}

The spontaneous emergence of spatiotemporal order is a prominent feature of physical systems driven far from equilibrium \cite{cross}. Optical systems provide valuable platforms for studying spatial self-organization, where the self-sustained patterns are typically encoded in the internal excitations of a nonlinear medium \cite{scrog, arecchi1}.
Besides their fundamental interest, optical spatial structures offer potential applications in information processing such as optical memories and registers \cite{dsolitons}. 

Recent experiments have provided paradigmatic examples of spatial self-organization in transverse nonlinear optics with cold atoms such as density, electronic and magnetic spatial ordering \cite{labeyrie1, camara, labeyrie2}. In the first case, the bunching of atoms due to optomechanical forces can provide positive feedback and lead to spatial instabilities \cite{ritsch1}. Other mechanisms for pattern formation in cold atoms rely instead on optical pumping and internal state nonlinearities.
In the regime of quantum degeneracy, the non-equilibrium phase transition corresponding to the onset of spatial self-organization has been interpreted as a quantum phase transition \cite{esslinger, gordon1}. Multimodal configurations have been shown to enrich significantly the physical scenario, allowing the realization of frustrated interactions and supersolid states with atomic gases  \cite{lev, esslinger2}.

The introduction of orbital angular momentum of light (OAM) and its relative ease of generation and control paved the way to a plethora of applications in optical manipulation and information technologies \cite{allen1, alison1}. Focusing on cold atoms, the use of OAM modes enables transfer of angular momentum to both motional and internal atomic degrees of freedom \cite{sonja1}. Macroscopic rotation of cold atomic gases is also achieved in the dispersive regime by means of rotating ring lattice trapping potentials \cite{sonja2, aidan}. Such geometries and related configurations, when applied to cold atoms, offer the possibility to engineer classical or quantum transport \cite{hanggi1} and to simulate topological condensed matter effects such as fractional quantum Hall physics \cite{hall}.

In this work we extend previous studies of optomechanical instabilities in cold atoms, allowing the input field to have a spatially dependent phase structure and carrying non-zero OAM. Externally controlled phase gradients are known to induce drifting pattern dynamics, as demonstrated with hot atomic vapors \cite{hot1, hot2}, and can be applied to control the motion of dissipative solitons \cite{OBH, OBH1, cornelia1}. Moreover, drifting dynamics induced by OAM was observed for patterned states in a single mirror system with photorefractive nonlinearity \cite{sciamanna1}. For a cold atomic gas with optomechanical nonlinearity, we show that such dynamics, corresponding to the motion of coupled structures of light and atoms, is capable of inducing controllable atomic mass transport. Our atomic ensemble is assumed to be kept at constant temperature with optical molasses, providing strong damping to the center of mass dynamics. Moreover, dealing with density redistribution-induced interactions, atoms behave as linear scatterers.
This opens the possibility of extending the present results to soft matter systems such as suspensions of colloidal particles \cite{ashkin,rogsari,dholakia}.  

The paper is organised as follows. In Sec. \ref{sec:level2} we review the known features of optomechanical self-structuring in a ring cavity for a spatially homogeneous input pump. In Sec. \ref{sec:level3} we discuss drifting pattern dynamics induced by OAM in the pump profile, which leads to the formation of self-organized rotating ring lattices. Finally, in Sec. \ref{sec:level4}, we compare our predictions with numerical results from particle dynamics simulations, showing the presence of macroscopic optomechanical transport corresponding to the pattern rotation.

\section{\label{sec:level2}The model}

In this section we discuss a theoretical model, adapted from Refs. \cite{saffman,tesio1}, describing the transverse dynamics of the cavity field and the density of the atomic ensemble. A linear stability analysis provides the threshold condition for transverse optomechanical modulation instabilities (MI) of the homogeneous (flat) stationary states.

\subsection{Model Equations}

\begin{figure}\begin{center}
\begin{tikzpicture}[scale = 2.2,
>=stealth, 
    axis/.style={thick,->},
    axisz/.style={dashed,->},
    wave/.style={thick,color=#1,smooth},
    polaroid/.style={fill=black!60!white, opacity=0.3}]

\definecolor{clr}{RGB}{110,110,110}

\shade[ball color = blue, opacity = 0.6] (1.5,0+.22) circle [radius = .28cm];

\textbf{\draw [->, ultra thick, purple] (+1+0+.22,1.7-.22) -- (+1.5+0+.22,1.7-.22);
}
\textbf{\draw [->, ultra thick, orange] (-.8+0+.22,1.7-.22) -- (-.15+0+.3,1.7-.22);
}
\textbf{\draw [->, ultra thick, blue] (-.01+.22,1.49) -- (-0.3,1.77);
}
\draw (3,1) node at (.18, 1.3) [text width=2.5cm] {$A_{I}(\mathbf{r})$};
\draw (3,1) node at (1.9, 1.65) [text width=2.5cm] {$E(\mathbf{r},t)$};
\draw (3,1) node at (1, 1.8) [text width=2.5cm] {$\tau$};
\draw (3,1) node at (1.9, -0.2) [text width=2.5cm] {$l,N_{0},T$};
    
\draw [fill=gray,ultra thin, cm={cos(25) ,-sin(25) ,sin(25) ,cos(25) ,(.3 cm,0.11 cm)}] (0,.35) to (.35,0) to (.35+.05 ,0+.05) to [out=150,in=-60] (.05+0,.05+.35) to (0,.35);
\draw [fill=gray,ultra thin, cm={cos(-25) ,-sin(-25) ,sin(-25) ,cos(-25) ,(0.04 cm,-1.16 cm)}] (3-.35,0) to (3,0+.35) to (3-.05,0+.35+0.05) to [out=-120,in=30] (3-.35-.05,0+.05) to (3-.35,0)  ;
\draw [fill=gray,ultra thin, cm={cos(25) ,-sin(25) ,sin(25) ,cos(25) ,(-.36 cm,1.32 cm)}] (3,1.7-.35) to (3-.35,1.7) to (3-.35-.05,1.7-.05) to (3-.05,1.7-.35-.05) to (3,1.7-.35) ;
\draw [fill=lightgray,ultra thin, cm={cos(-25) ,-sin(-25) ,sin(-25) ,cos(-25) ,(0.64 cm,0.045 cm)}] (0+.35,1.7) to (0,1.7-.35) to (0+.05,1.7-.35-.05) to (0+.35+.05,1.7-.05) to (0+.35,1.7) ;

\textbf{\draw [->, thick,dotted, purple] (0+.22,1.7-.22) -- (2.55+.22,1.7-.22) to (.55, .21) to (.55 +1.96, .21) to (0+.22,1.7-.22)  ;
}


\end{tikzpicture}\caption{\small Sketch of our cavity setup. A  structured beam with amplitude $A_{I}(\mathbf{r})$ and phase $\exp{(il\varphi)}$ drives a single longitudinal cavity mode. The ring cavity has effective length $L$ and mirror transmittivity $\tau$. The intra-cavity field $E(\mathbf{r},t)$ interacts with a cloud of two-level atoms of thickness $l$, average atomic density $N_{0}$ and  temperature $T$.}
\label{setup}
\end{center}\end{figure}
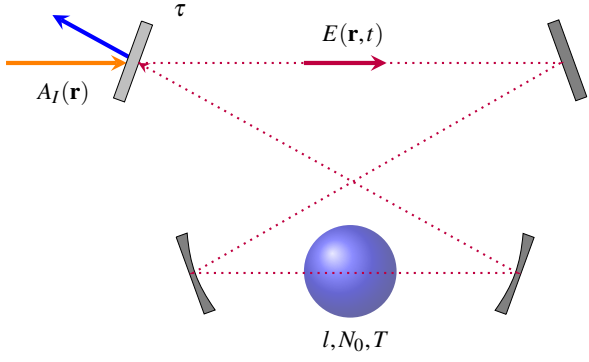

We consider a cold thermal cloud of two-level atoms within a planar ring cavity geometry of effective length $L$, sketched in Fig.$\!$  (\ref{setup}). 
Theoretical models describing the coupled dynamics of cavity field and density modulations of a cold atomic gas involve the paraxial wave equation for a single longitudinal mode of the cavity field $E(\mathbf{r},t)$ coupled to the atomic density distribution $N(\mathbf{r},t) = N_0\, n(\mathbf{r},t)$, where $N_0$ is the average density of the sample and $\mathbf{r}$ denotes the transverse spatial coordinate \cite{tesio1}. The field equation is derived in the slowly varying envelope, rotating wave and mean field approximations as follows \cite{lugiato1}:

\begin{equation}
\frac{\partial E}{\partial t}=-\kappa\left\{(1+i\theta)E + A_{I}(\mathbf{r}) - \frac{\tilde{\gamma}}{1+|E|^2}\, n\, E + ia\nabla_{\perp}^2 E\right\} 
\label{field}
\end{equation}

where $\theta$ is the cavity-pump detuning, describing the linear shift on the cavity field, $A_I(\mathbf{r})$ represents a spatially dependent pump and diffraction is described by the transverse Laplacian $\nabla_{\perp}^2 $. Here we focus on a pump profile of the form $A_I(\mathbf{r}) = A_I(r)\exp{(il\varphi)} $, where $A_I(r)$ is a generic radial profile and $r, \varphi$ are the radial and azimuthal coordinates respectively. In addition, $\kappa = c\tau/L $ represents the cavity linewidth and $\sqrt{a}=\sqrt{\lambda L /4\pi \tau}$ the diffraction length, where $\lambda$ is the light wavelength, $\tau$ the mirror trasmittivity and $c$ the speed of light in vacuum. The response of the two level saturable medium\footnote{The characteristic dependence is obtained by adiabatic elimination of the atomic variables.} is parametrized in terms of the complex susceptibility  $\tilde{\gamma} = 2C(1+i\Delta)$, with $\Delta = 2\delta/\Gamma$, where $\delta$ is the light detuning with respect to the atomic resonance. Finally, the parameter $C = b_0/ 2\tau(1+\Delta^2)$ is known as cooperativity parameter and it depends on the optical thickness of the sample at resonance $b_0$, $\tau$ and $\Delta$.
For our description of atomic motion, we focus on the limit when the population of the excited atomic state is negligible. This is captured by the saturation parameter\footnote{The cavity field $E$ is rescaled by the saturation intensity at detuning $\Delta$, i.e. $[I_\textrm{sat}(1+\Delta^2)]^{\frac{1}{2}}$, where $I_\textrm{sat}$ represents the saturation intensity at resonance.} $s=|E(\mathbf{r},t)|^2$. In general, the resulting dipole force $\mathbf{f}_{\textrm{dip}}$ acting on the atom center of mass can be derived from the AC Stark potential:

\begin{equation}
\mathbf{f}_{\textrm{dip}} = -\nabla_{\perp}U_{\textrm{dip}}(\mathbf{r},t) = -\frac{\hbar\Gamma\Delta}{4}\nabla_{\perp}\left(\frac{s(\mathbf{r},t)}{1+s(\mathbf{r},t)}\right).
\label{force}
\end{equation}

Note the dependence of Eq. (\ref{force}) on the detuning $\Delta$, such that the atom is trapped in high (low) intensity regions when $\Delta<0$ ($\Delta>0$).  
Assuming the presence of optical molasses, we adopt a classical treatment of positions $\mathbf{r}_j$ and momenta $\mathbf{p}_j$ ($i,j=1,\dots,N$), leading to the canonical set of stochastic differential equations: 

\begin{equation}
\frac{d \mathbf{r}_{j}}{dt} = \frac{\mathbf{p}_j}{m}, \quad \frac{d \mathbf{p}_{j}}{dt} = \mathbf{f}_{\textrm{dip}}(\mathbf{r}_j,t)-\gamma \frac{\mathbf{p}_j}{m} + \xi_j(t)
\label{stoch}
\end{equation}

where  $m$ is the atomic mass, $\gamma$  describes friction (momentum damping) and  $\mathbf{\xi}_j(t)$ is a stochastic variable such that $\langle\mathbf{\xi}_j(t)\rangle = 0$ and $\langle\mathbf{\xi}_j(t)\mathbf{\xi}_j'(t')\rangle=2D_p\delta_{jj'}\delta(t-t')$, with $D_{p} $ being the diffusion constant in the space of momenta. In the thermodynamic limit ($N\rightarrow\infty$), the above equations map to the Chandrasekhar equation for the phase space distribution function $f(\mathbf{r},\mathbf{p},t)$ in the 1-particle picture \cite{chandra, rogsari}: 

\begin{equation}
\frac{\partial}{\partial t} f = \frac{\mathbf{p}}{m}\cdot \nabla_{\perp} f + \mathbf{f}_{\textrm{dip}}(\mathbf{r},t)\cdot \nabla_{\mathbf{p}}f + \gamma \left[\nabla_{\mathbf{p}}(f\mathbf{p})+\frac{m}{\beta}\nabla^2_{\mathbf{p}}f\right]
\label{chandrasekhar}
\end{equation}

where $\beta = 1/k_B T$. Note that, in the presence of damping, the collisional contribution in Eq. (\ref{chandrasekhar}) is replaced by the term $\nabla_{\mathbf{p}}(f\mathbf{p})+\frac{m}{\beta}\nabla^2_{\mathbf{p}}f$, describing relaxation in momentum space. In the limit of strong viscous damping (more generally, when $t\gg 1/\gamma$), the momentum $\mathbf{p}$ is eliminated from the dynamics and one derives the Smoluchowski equation for the density distribution $n(\mathbf{r},t)$, by direct integration of Eq. (\ref{chandrasekhar}) in momentum space, namely: 

\begin{equation}
\frac{\partial n}{\partial t}= \beta D\nabla_{\perp}\cdot\left[n\,\mathbf{f}_{\textrm{dip}}\right]+D\nabla_{\perp}^{2}n
\label{smolu}
\end{equation}

where diffusive dynamics, as an effect of the optical molasses, is characterized by the constant $D=1/\gamma m\beta$ \cite{molasses}. Clearly, in the case of hot atoms ($\beta \rightarrow 0$), the diffusive term dominates and one has $n(\mathbf{r},t)\approx 1$, i.e. the homogeneous state.  Generally speaking, MI in our system is due to a competition between two-level and density-driven nonlinearities. In what follows, we perform a linear stability analysis, showing that density inhomogeneities alone are sufficient to achieve (optomechanical) MI. The features of the resulting self-organized states in the presence of OAM are discussed in Sec. (\ref{sec:level3}). 

\subsection{Linear Stability Analysis}

The family of homogeneous stationary states of our system is simply found imposing that all derivatives in both Eqs. (\ref{field}, \ref{smolu}) vanish. For the cavity field, i.$\!$ e. $\partial E_{0}/\partial t = 0$ and $\nabla^{2}_{\perp}E_{0} = 0$ in Eq. (\ref{field}), this yields:

\begin{equation}
A_{I} = \left[1+i\theta + \frac{\tilde{\gamma}}{1+|E_{0}|^2}\right]\, E_{0}
\label{bistable}
\end{equation}

with $n_0 =1$. A consequence of the two-level nonlinear term in Eq. (\ref{bistable}) is that the homogeneous field intensity $|E_0|^2$, as a function of the pump intensity $A_I$, is not necessarily single valued \cite{lugiato1}. It is easy to see that, in the low saturation limit, our model Eqs. (\ref{field}) and (\ref{smolu}) read as follows: 

\begin{equation}
\frac{\partial E}{\partial t'}=-(1+i\theta)E + A_{I}(\mathbf{r}') -  2iC\Delta\, n\, E + i\nabla_{\perp}^2 E 
\label{fieldlow}
\end{equation}

\begin{equation}
\frac{\partial n}{\partial t'}= \sigma D'\nabla_{\perp}\cdot\left[n\nabla_{\perp}|E|^2\right]+D'\nabla_{\perp}^{2}n  
\label{smolu1}
\end{equation}

where time and space coordinates are rescaled as $t'=\kappa t$, $\mathbf{r}'= \mathbf{r}/\sqrt{a}$, $D'=D/\kappa a$, while the constant $\sigma = \hbar\Gamma\Delta/4k_{B}T$ denotes an optomechanical coupling strength. Moreover, Eq. (\ref{bistable}) now simply reads $A_{I} = \left[1+i(\theta + 2C\Delta) \right]\, E_{0}$, so that there is always a one-to-one correspondence between the values of $|A_{I}|^2$ and $|E_{0}|^2$ in the low saturation limit. Perturbations around steady states values $E_0$ and $n_0$ are introduced in Fourier space as $\delta E(\mathbf{q},t') = a\, e^{i(\mathbf{q}\cdot\mathbf{r}'+\nu t')}+b^*\, e^{-i(\mathbf{q}\cdot\mathbf{r}'+\nu^* t')}$, $ \delta n(\mathbf{q},t') = c\, e^{i(\mathbf{q}\cdot\mathbf{r}'+\nu t')}+c^*\, e^{-i(\mathbf{q}\cdot\mathbf{r}'+\nu^* t')}$. Thus, defining the variable $\mathbf{x} = (a,b,c)^{\textrm{T}}$, the linearized system obtained from Eqs. (\ref{fieldlow}, \ref{smolu1}) can be cast in the usual matrix form $(\mathbf{M}-\nu\mathbb{I})\mathbf{x}=\mathbf{0}$, where $\mathbf{M}$ is the following matrix:  

\small
\begin{equation}\begin{split}
&\mathbf{M} = \\
&i^{-1}\left(
\begin{array}{ccc}
-1 -i(\mathbf{q}^2+2 C \Delta + \theta) & 0 & -2 i C \Delta  E_0 \\
 0 & -1+ i(\mathbf{q}^2+2 C  \Delta + \theta) & 2 i C \Delta  E_0^* \\
 -D' \mathbf{q}^2 \sigma  E_0* & -D' \mathbf{q}^2 \sigma  E_0 & -D' \mathbf{q}^2 \\
\end{array}
\right).
\end{split}
\label{matrix}
\end{equation}
\normalsize

\begin{figure}

\includegraphics[scale=0.66]{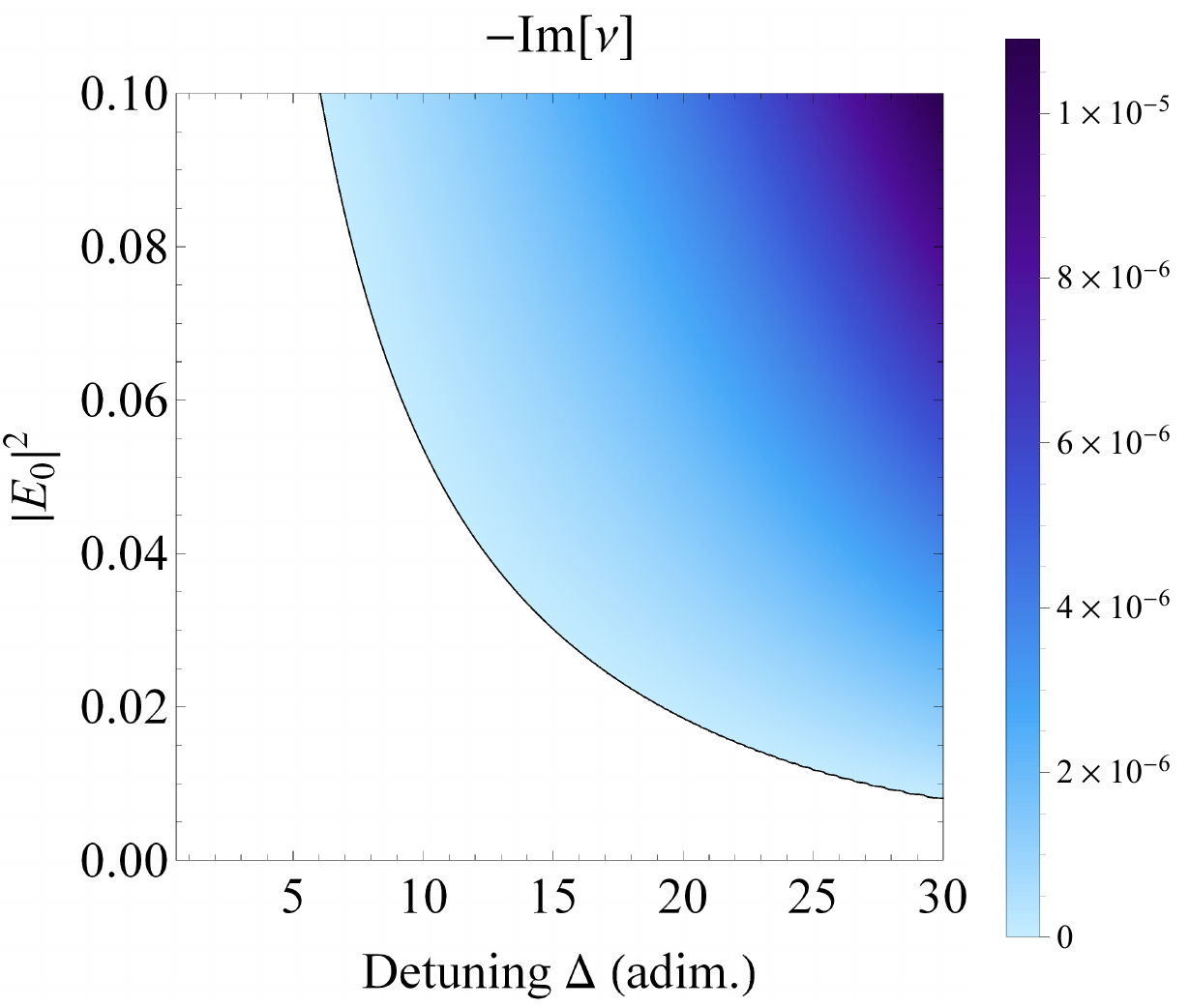}
\includegraphics[scale=0.66]{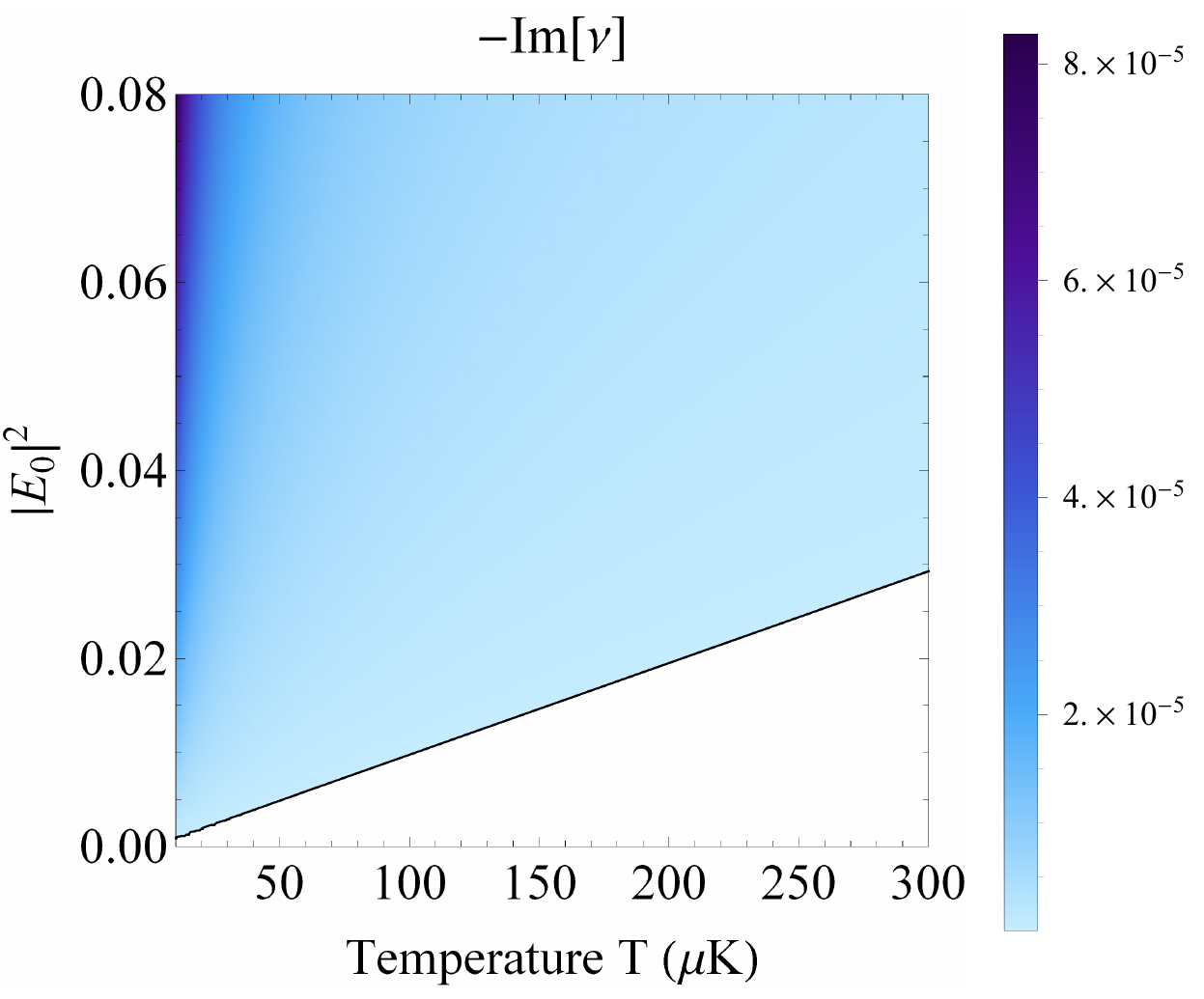}

\caption{\small 2D diagrams from linear stability analysis showing the growth rate $-\textrm{Im}[\nu]$ dependence in two parameter spaces. (Top) GR as a function of $|E_0|^2$ and $\Delta$, sufficiently far from the atomic resonance (low saturation) at $\sigma=12$. (Bottom) GR as a function of $|E_0|^2$ and temperature $T$ at $|\Delta| = 20$.
Parameters are chosen as follows: $\theta=-100$, $D'=10^{-8}$ and $\mathbf{q}^2=\mathbf{q}^2_{c}$. Instability regions are obtained above the threshold in Eq. (\ref{threshold}), represented by the solid black line.}
\label{fig:fig}
\end{figure}

By imposing the marginal stability condition, i.e. $\nu = 0$ or  $\det(\mathbf{M}) = 0$, one finds the following analytical threshold:

\begin{equation}
|E_{0}|^2 = \frac{1+(\mathbf{q}^2 + \theta + 2C\Delta)^2}{4C\Delta\sigma(\mathbf{q}^2+ \theta + 2C\Delta)}
\label{threshold}
\end{equation}

leading to the critical wavevector $\mathbf{q}^2_c = 1- (\theta + 2C\Delta)$. 
Here we study the linear stability properties by spanning two different parameters regions. The two instability diagrams in Fig. (\ref{fig:fig}) show the dependence of the linear growth rate (GR), given by -$\textrm{Im}[\nu]$, as a function of $|E_{0}|^2 $ vs. detuning $\Delta$ (Fig.\ref{fig:fig} - Top)\footnote{Only the case $\Delta>0$ is shown as we have symmetric behaviour for $\Delta<0$.} and $|E_{0}|^2$ vs. temperature $T$ (Fig.\ref{fig:fig} - Bottom). The value of  $\textrm{Im}[\nu]$ is computed by estimating the minimum imaginary part among the eigenvalues of $\textbf{M}$ in Eq. (\ref{matrix}). Finally, the homogeneously pumped system spontaneously selects an hexagonal patterned state for values of $|E_{0}|$ close to threshold \cite{hex}. We remark here that, assuming low saturation, purely optomechanical structures occurs in the proximity of the threshold only since, for stronger pumping, electronic nonlinearity becomes relevant.

\section{\label{sec:level3}Pattern dynamics with structured phase}

In this section we analyze the effect of the spatially dependent pump $A_{I}(\mathbf{r}')$ on the transverse geometry and dynamics of the patterned states of our system. Before presenting numerical results, we show how drifting pattern dynamics arises from Eqs. (\ref{fieldlow}, \ref{smolu1}), based on the argument from Refs. \cite{OBH,Kerr}.

\subsection{Drifting pattern dynamics}

In order to include OAM in our system, the pump rate $A_{I}(\mathbf{r}')$  must have an azimuthally dependent phase $\exp(il\varphi)$, where $l\in\mathbb{Z}$. Indeed, the factor $\exp{(il\varphi)}$ generates $l$ intertwined phase fronts and a non-trivial topological structure due to the phase singularity at $\mathbf{r}'=0$ \cite{alison1}.  In general, any phase profile including the factor $\exp{(il\varphi)}$ is responsible for a non-vanishing contribution of OAM through the transverse plane.
Let us further simplify the analysis recalling that the stationary solution of Eq. (\ref{smolu1})  is given by the Gibbs distribution \cite{saffman}: 

\begin{equation}
n_{eq}(\mathbf{r}') = \frac{\exp[\beta U_{\textrm{dip}}(\mathbf{r}')]}{\int_{V}\exp[\beta U_{\textrm{dip}}(\mathbf{r}')]}=\frac{\exp[\sigma |E(\mathbf{r}')|^2]}{\int_{V}\exp[\sigma |E(\mathbf{r}')|^2]}.
\label{gibbs}
\end{equation} 

The steady states of the coupled system above MI can in principle be obtained by integrating numerically Eq. (\ref{fieldlow}), simply eliminating the density distribution. Thus, Eq. (\ref{gibbs}) acts back as the nonlinear term in the field equation, namely \cite{tesio1}:  

\begin{equation}
\frac{\partial E}{\partial t'}=-(1+i\theta)E + A_{I}(\mathbf{r}') -  2i\,C\Delta\,\, n_{eq}(|E|^2)\, E + i\nabla_{\perp}^2 E 
\label{fieldlow1}
\end{equation}

where we emphasize here the dependence of $n_{eq}(\mathbf{r}')$ on the field intensity. Therefore, the coupled system of Eqs. (\ref{fieldlow}, \ref{smolu1}) reduces now to Eq. (\ref{fieldlow1}) only. Let us consider the local transformation $E(\mathbf{r}',t') = \tilde{E}(\mathbf{r}',t') \exp(il\varphi)$. The corresponding equation for $\tilde{E}(\mathbf{r}',t')$ reads now as follows:

\begin{equation}\begin{split}
\frac{\partial \tilde{E}}{\partial t'} +&\frac{2l}{r'} \nabla_{\perp} \tilde{E}=-\left(1+\frac{l^2}{r'^2} \right)\tilde{E}-i\left(\theta-\frac{l}{r'^2}\right)\tilde{E}+\\
&  A_{I}(r') - 2i\,C\Delta\,\, n_{eq}(|\tilde{E}|^2)\, \tilde{E} + i\nabla_{\perp}^2 \tilde{E}.
\label{gauge}
\end{split}\end{equation}

One can recognize that the LHS of Eq. (\ref{gauge}) has the form of a covariant derivative $D_{t'}=\frac{\partial}{\partial t'} + \mathbf{v}_{\textrm{dr}}\cdot\nabla_{\perp}$, i.e. the expression of a time derivative in a locally co-moving reference frame, according to the velocity $\mathbf{v}_{\textrm{dr}}(\mathbf{r}') = 2l\nabla_{\perp}\varphi  = \frac{2l}{r'} \hat{\varphi}$. Thus, in the presence of OAM, the family of patterned solutions are seen in the laboratory frame drifting with a velocity $\mathbf{v}_{\textrm{dr}}(\mathbf{r}')$ \cite{OBH}. Furthermore, such an argument carries implications also for the geometry of the patterned state in our case. Indeed, the RHS of Eq. (\ref{gauge}) shows a spatially dependent contribution to the decay rate and cavity detuning. Therefore, the spatial rigidity of the patterned phases is broken according to the spatial symmetry of the pump $A_{I}(\mathbf{r}')$ and the stationary solution is retrieved in a rotating reference frame defined (in polar coordinates) by $\tilde{\varphi}(r',t') = \varphi - \omega(r') t'$, with the angular frequency  $\omega(r')=2l/r'^{\,2}$. This corresponds to a differential rotation of the concentric patterned solutions at a fixed radius, as found in Ref. \cite{Kerr}. In what follows, we support the above observation by integrating Eq. (\ref{fieldlow1}) numerically. 

\subsection{Numerical Results}

We shall restrict here our analysis to a 2D radial "top hat" profile with rapidly vanishing tails. A commonly used example is given by the following: 

\begin{equation}
A_{I}(\mathbf{r}')= \frac{A_{I}}{2}\left\{1-\textrm{tanh}[\eta(r'-r'_{0})]\right\}\exp{(il\varphi)}
\label{tophat}
\end{equation}

where the constant $\eta>0$ determines the side steepness of the flat part with radius $r'_0$. The advantage of using Eq. (\ref{tophat}) is twofold: the used boundary conditions do not affect numerical results and allow us to investigate the transverse 2D dynamics induced by the structured phase factor $\exp{(il\varphi)}$ on a wide area inside the integration domain.

\begin{figure}
\includegraphics[scale=0.22]{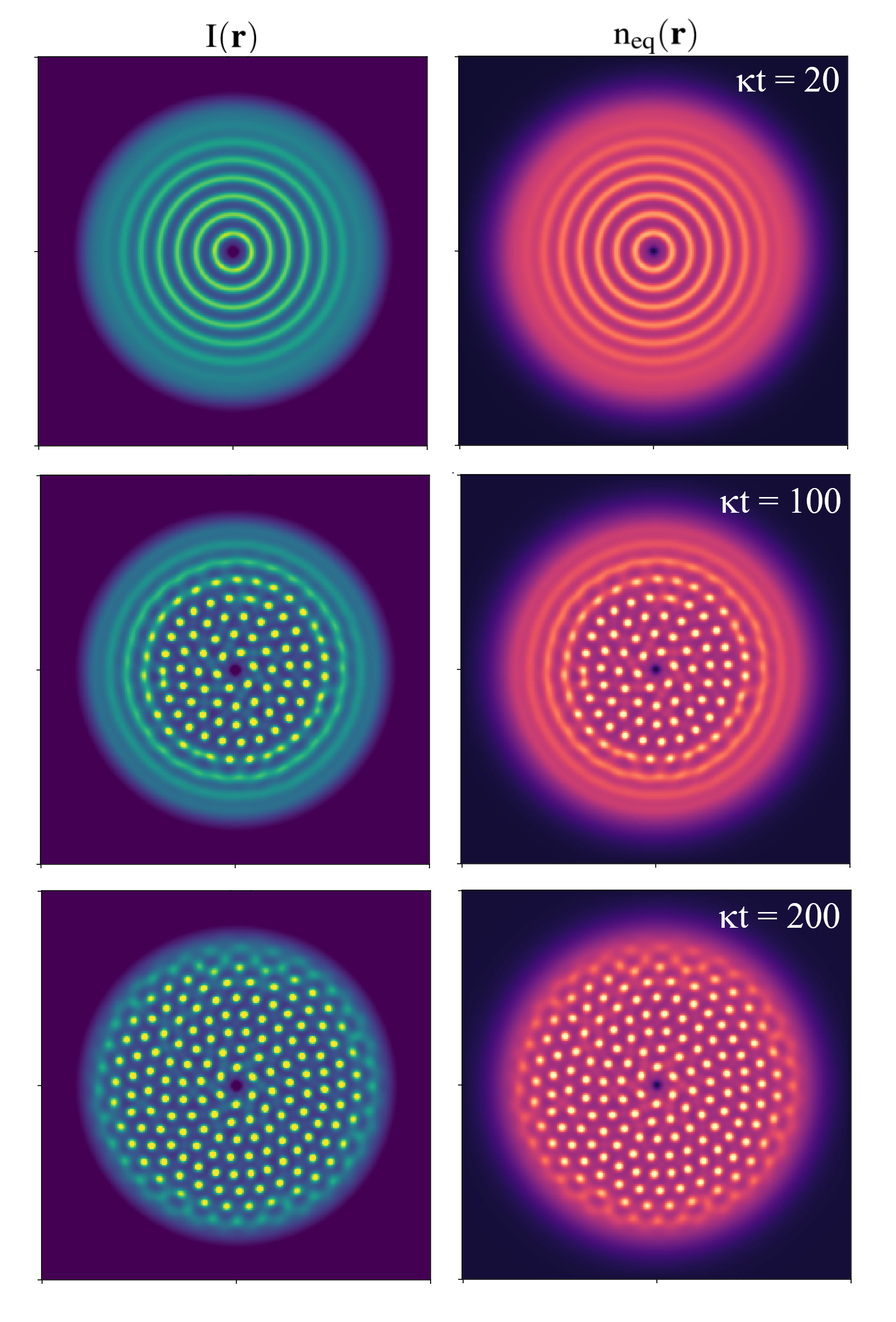}
\caption{\small Formation of self-organized ring lattices with a red detuned pump intensity $5\%$ above threshold and $l =1$. The 2D intensity $I(\mathbf{r})$ (left) and density $n_{eq}(\mathbf{r})$ are shown at different integration times. Parameters are chosen as follows: $\theta = 4,$ $\, C\Delta = -3.5$ $ \,(\Delta<0),$ $\, \sigma = -25$. Note that atoms bunch towards the maxima of intensity.}
\label{patterns}
\end{figure}

We integrate Eq. (\ref{fieldlow1}) using a Fourier split-step method. A domain of $10$ critical wavelengths is discretized in 256$\times$256 points and time step $dt' = 5\times 10^{-3}$. Fig. (\ref{patterns}) shows an example of optomechanical self-structuring with $\textrm{OAM}$ index $l=1$. The presence of a helical phase $\exp{(il\varphi)}$ in the pump profile (\ref{tophat}) causes the field to vanish at $\mathbf{r}'=0$ and induces the formation of diffractive rings whose intensity exceeds the MI threshold. Such rings are spaced by a critical wavelength, visible from Fig. (\ref{patterns}) when $t' =\kappa t =20$. 
Since the intensity between contiguous rings is below the MI threshold, each ring undergoes a 1D angular instability independently, leading to a set of concentric ring lattices structures \cite{Kerr}. Moreover, by varying $\eta$ and $r'_0$, one can also influence the features of the patterned states. For example, increasing the steepness with the parameter $\eta$ may lead to outer diffraction rings faster achieving azimuthal MI. For any balanced choice of such parameters, self-structuring dynamics always leads to a set of ring light-density lattices, each one rotating independently from the others. Similar behaviour is found with higher order OAM modes ($l>1$).

\begin{figure}
\includegraphics[scale=0.44]{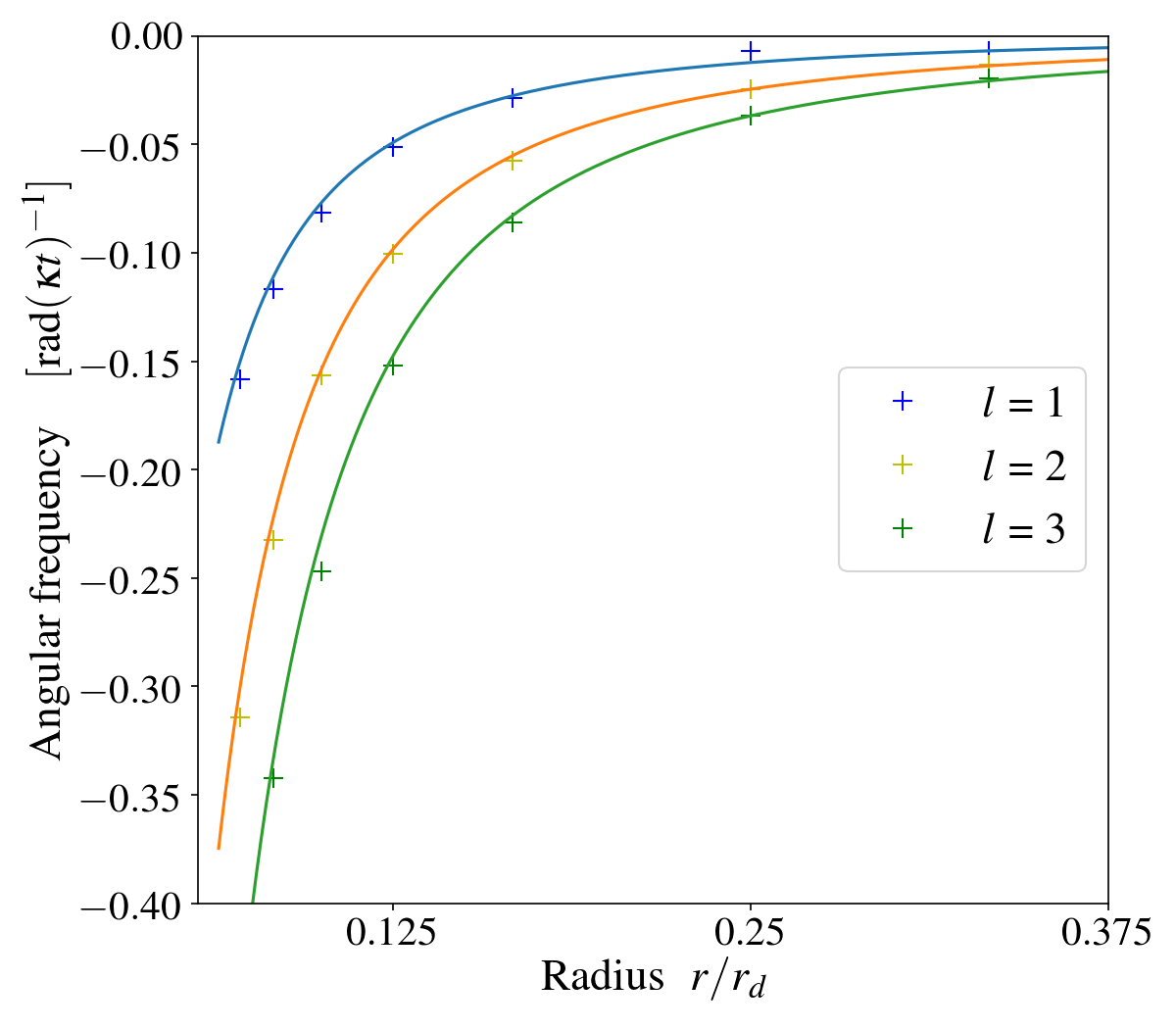}
\caption{\small Results from the 1D integration of Eq. (\ref{fieldlow1}). Values of the angular velocity of the patterned states are estimated at different radii (in units of half the domain size $r_d$). Each radius is chosen by simply adjusting the size of the integration domain. Solid lines represent the predicted angular frequencies $\omega$ for the three OAM cases $l=1,2,3$.}
\label{agr}
\end{figure}

The self-organized light-atom ring lattices display differential rotation, in accordance to the angular frequency $\omega(r')=2l/r'^{\,2}$. To show this, we consider the field in Eq. (\ref{fieldlow1}) at a fixed radius $r'=r'_0$, mapping the variation of $\varphi \in [0, 2\pi]$ to an effective 1D model with periodic boundary conditions and  an input pump of the form $A_{I}(\varphi) = A_{I}e^{il\varphi}$. The same argument is repeated in Sec. \ref{sec:level4} for studying atomic transport. Fig. (\ref{agr}) shows a comparison, for $l=1,2,3$, of the rotation frequency estimated at several radial distances and showing good agreement with the predicted $\omega(r')$. Analogous rotational behaviour can be expected for the case of bistable optomechanical dissipative solitons, as those found in \cite{optsolit}. OAM or other structured phase pump profiles can thus be used in this case to control the motion and the effective interactions of such localized dissipative structures \cite{OBH1, cornelia1}.

\begin{figure}
\includegraphics[scale=0.28]{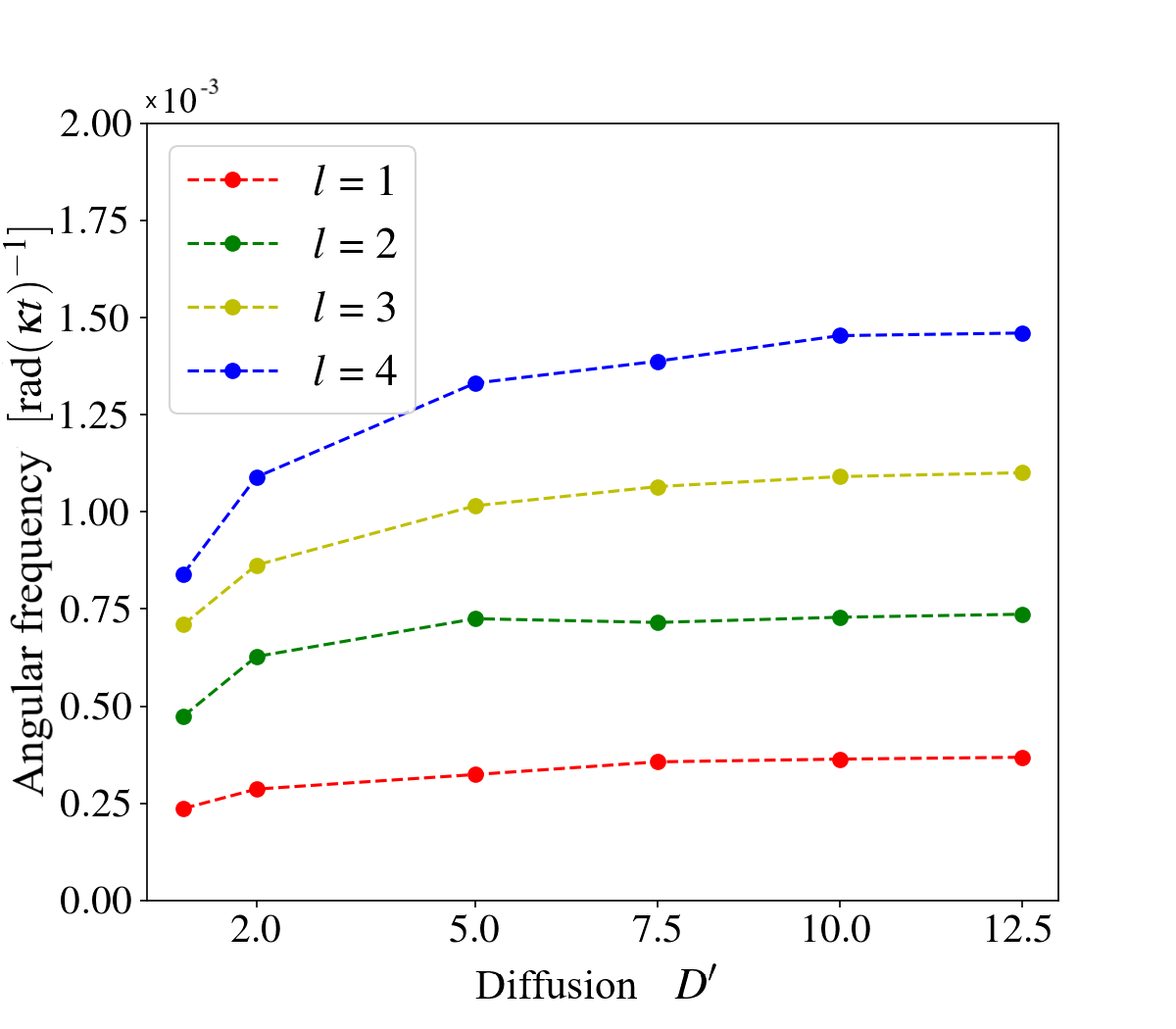}
\caption{\small Results from the 1D integration of the coupled model of Eqs. (\ref{fieldlow}, \ref{smolu1}). Estimated values of the pattern rotation frequency are plotted against the rescaled diffusion coefficient $D'$ (at fixed temperature $T$) for the cases $l=1,\dots,4$. 
As a result, faster diffusion processes imply higher values of experimentally measurable ring rotation frequency.}
\label{difcof}
\end{figure}

In the rest of this section, we discuss the effect of varying the diffusion coefficient $D$ on the ring lattice rotational dynamics.  In order to take into account this dependence and obtain realistic  values of the rotation frequency comparable with experiments, we numerically integrate the coupled cavity Eq. (\ref{fieldlow}) and the Smoluchowski Eq. (\ref{smolu1}), for the 1D azimuthal case at fixed radius\footnote{By means of a second order Crank-Nicolson scheme for coupled equations.}. Note that, since we have $D=1/\gamma m\beta$, any variation of $D$ at constant temperature is inversely proportional to a variation in the momentum damping coefficient $\gamma$, in the particle Eqs. (\ref{stoch}). Results are shown in Fig. (\ref{difcof}), where values of the estimated angular frequency are plotted against the rescaled diffusion coefficient $D'=D/\kappa a$. For higher values of $D'$ (decreasing the friction $\gamma$), one observes a faster ring lattice rotation speed which saturates to a value roughly proportional to the index $l$. 
On the other hand, by increasing the friction $\gamma$, the rotation of the ring lattices created by the dipole interaction between atoms and light is affected by a drag, due to slower atomic diffusion. In the limit of very small diffusion ($D'\ll 1$), this drag can compensate the effect of the phase gradient introduced by the pump OAM and the concentric ring lattices would be expected to remain stationary. 
The above observations have profound consequences from the point of view of atomic center-of-mass dynamics since drifting self-organized light-density modulations induce transport of the self-trapped atoms, as  shown in the next section.

\section{\label{sec:level4}Atomic transport}

In this section we demonstrate OAM induced atomic transport corresponding to the rotation of the self-organized ring lattices. This is done by simply addressing the average momentum of the atomic distribution which quantifies the macroscopic rotational motion of the atoms, occurring at the onset of self-organization.  We restrict ourselves to the 1D case i.e. atoms  confined in a ring geometry at a fixed radius within the cavity 2D transverse domain. Such a case can be realized experimentally by appropriately using Laguerre-Gauss beams of a given OAM  in order to excite a single patterned ring \cite{sonja1}.
 
\subsection{Average momentum}

We start our observations by recalling that, at equilibrium, the Chandrasekhar Eq. (\ref{chandrasekhar}) for a dilute 1D atomic gas at constant temperature $T$ is solved by the Maxwell-Boltzmann distribution $\bar{f}(p)$, namely  \cite{huang}: 

\begin{equation}
\bar{f}(p) = \left(\frac{\beta}{2\pi m}\right)^{3/2} \exp\left[-\frac{\beta(p-\langle p\rangle)^2}{2m}\right]
\label{maxwboltz}
\end{equation}

where  $\langle p\rangle$ is the average (bulk) momentum around which the distribution $\bar{f}(p)$ is centered.  Since we are interested in the strong friction limit, we assume that the total  phase space distribution has the following factorized form \cite{saffman, huang}: 

\begin{equation}
f(r,p,t) = \bar{f}(p) \; n(r,t)
\label{findist}
\end{equation}

where $r$ varies over a finite interval, depending on the chosen radius. Macroscopic transport in this regime can be addressed by looking at the average momentum $\langle p \rangle$, whose time evolution is obtained multiplying Eq. (\ref{chandrasekhar}) by the momentum $p$ and integrating over momentum space \cite{davies}:

\begin{equation}\begin{split}
\frac{\partial n \langle p \rangle}{\partial t} = \frac{1}{m} \int p^2  \,\,\frac{\partial f}{\partial r}\,dp  + \textrm{f}_{\textrm{dip}} \int p \,\frac{\partial f}{\partial p}\, dp \\
+\gamma \int p \,\, \frac{\partial}{\partial p} (p f)\,dp + \frac{\gamma m}{\beta} \int p\,\,  \frac{\partial^2 f}{\partial p^2}\, dp
\label{moment1}
\end{split}\end{equation}

where $\textrm{f}_{\textrm{dip}}$ is the dipole force in Eq. (\ref{force}), expressed in the low saturation limit. We now rearrange Eq. (\ref{moment1}), according to the following simple manipulations. Firstly, we have that $\int p^2  \,\,\frac{\partial f}{\partial r}\,dp = \frac{\partial}{\partial r} n \langle p^2\rangle$, where $\langle p^2\rangle = (m\beta)^{-1}$ is the second moment of $\bar{f}(p)$ in Eq. (\ref{maxwboltz}). Moreover, $\int p \,\frac{\partial f}{\partial p}\, dp = -n$, $\int p \,\, \frac{\partial}{\partial p} (p f)\,dp = - n \langle p\rangle$ and $ \int p\,\,  \frac{\partial^2 f}{\partial p^2}\, dp = 0$, by simple calculus arguments. Thus, at equilibrium, Eq. (\ref{moment1}) reads: 

\begin{equation}
0 = \frac{1}{m^2\beta} \frac{\partial}{\partial r} n(r,t)  -\left[\textrm{f}_{\textrm{dip}}(r,t) +\gamma  \langle p \rangle\right] n(r,t).
\label{momdyna}
\end{equation}

Atomic currents can thus be obtained integrating Eq. (\ref{momdyna}) along the 1D azimuthal domain. However, the first term in its RHS vanishes in our case\footnote{Since the density $n(r,t)$ is always a periodic function in our azimuthal domain at fixed radius.} so that, after integration, the average momentum finally reads:

\begin{equation}\begin{split}
\langle p(t) \rangle _{\textrm{density}} =- \frac{1}{\gamma m}\int \textrm{f}_{\textrm{dip}}(r,t) n(r,t)\,dr\\
=- D\beta \int \textrm{f}_{\textrm{dip}}(r,t)n(r,t)\,dr.
\label{finmom}
\end{split}\end{equation}

 This quantifies the steady state mass current along a rotating ring in the overdamped limit, from the knowledge of $n(r,t)$ only.  In what follows, we compare $\langle p(t) \rangle _{\textrm{density}}$ with the ensemble averages obtained from simulating particle dynamics.

\subsection{Particle model}

Since a critical wavenumber $q_c$ is selected at the MI threshold of Eq. (\ref{fieldlow}), it is convenient to re-define here atomic position $\bar{r}_j = q_c r_j$ and momenta $\bar{p} = p_j/\hbar q_c$, so that the set of particle Eqs. (\ref{stoch}) in 1D now reads \cite{carl}:

\begin{equation}
\frac{d \bar{r}_j}{d t'} = 2 \omega_p  \bar{p}_j, \quad \frac{d \bar{p}_j}{d t'} = -\frac{\Gamma \Delta}{4 \kappa}\frac{\partial}{\partial \bar{r}} |E(\bar{r}=\bar{r}_j,t')|^2 -\bar{\gamma}\, \bar{p}_j + \bar{\xi}_j(t')
\label{stoch1}
\end{equation}

where $\omega_p = \hbar q_c/2m\kappa$ defines characteristic energy scales of an atom oscillating in a standing wave potential of modulation $q_c$. In the overdamped case, one imposes momenta $\bar{p}_j$ at their steady state values, leading to the following epression of the average momentum or \textit{atomic current}:

\begin{equation}
\langle \bar{p}_{j}(t') \rangle_{\textrm{particles}} = -\frac{\Gamma \Delta}{4 \kappa\bar{\gamma}}\,\,\left\langle\frac{\partial}{\partial \bar{r}} \left|E(\bar{r}=\bar{r}_j,t')\right|^2 \right\rangle
\label{finmompart}
\end{equation}

\begin{figure}
\includegraphics[scale=0.27]{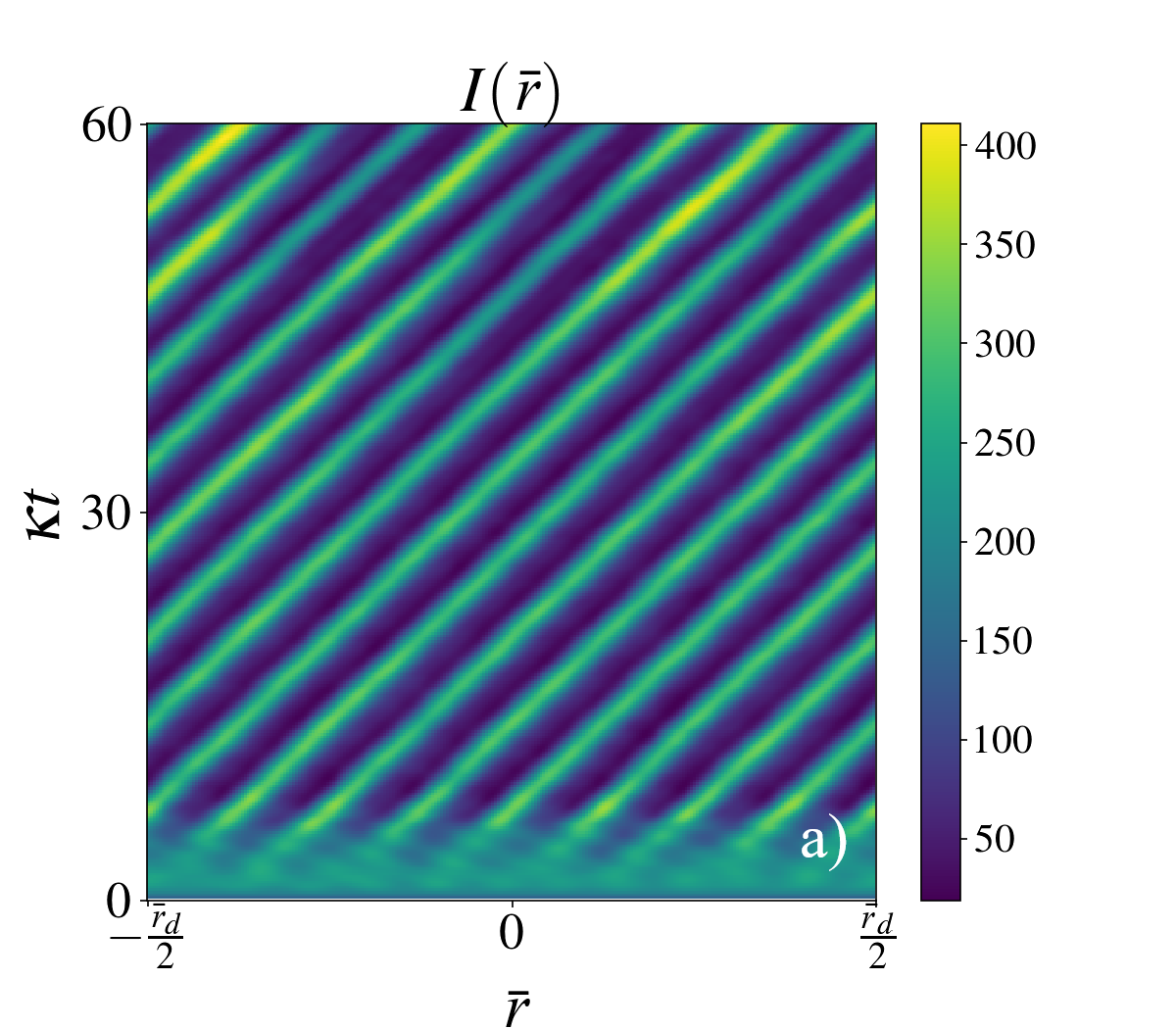}
\includegraphics[scale=0.27]{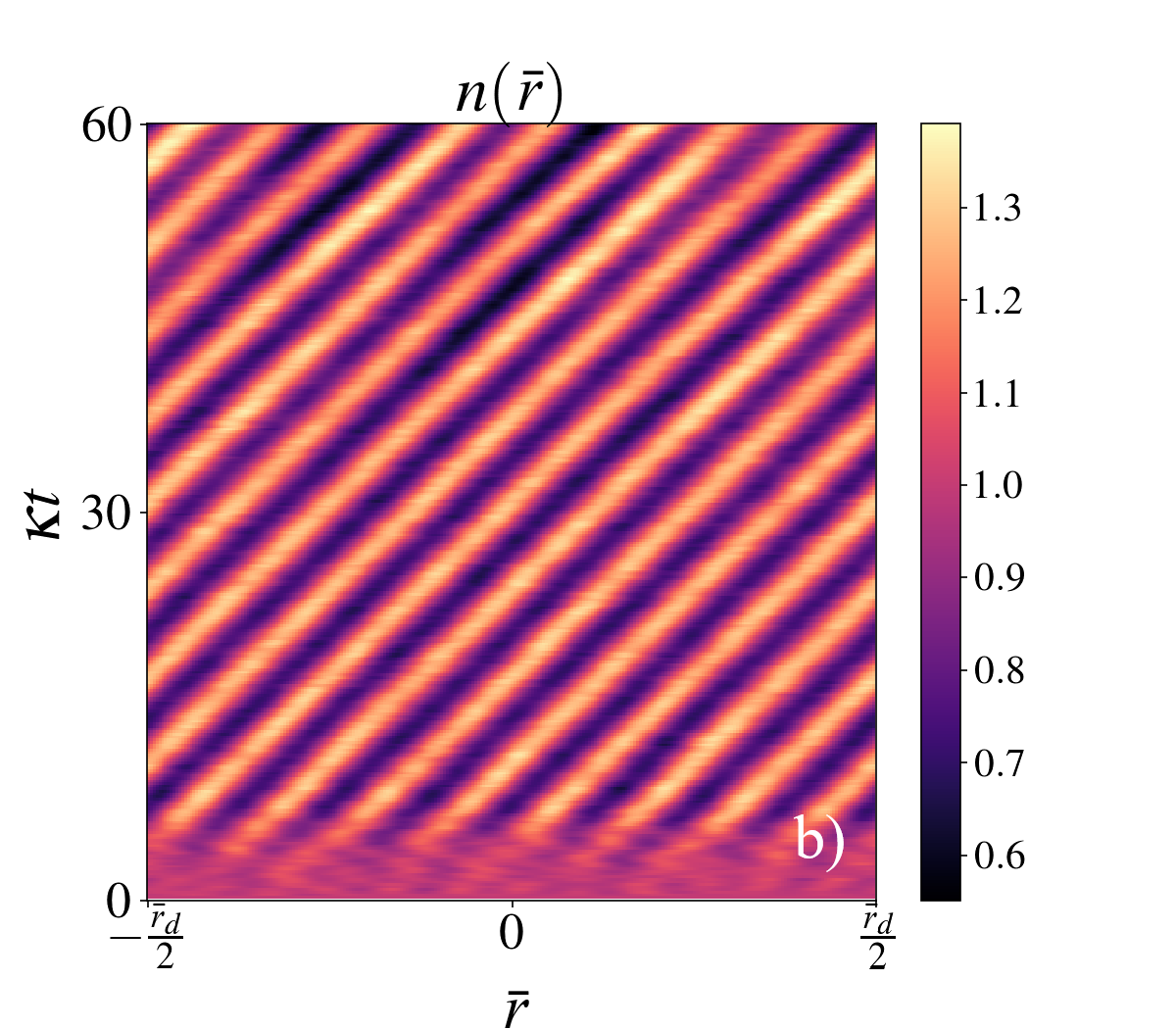}
\caption{\small Numerical results from the 1D particle simulation for a set of  $5\times10^4$ atoms. (a) $\kappa t$ vs. $\bar{r}$ plot showing the time evolution ($\kappa t_{max} = 60$) of the 1D cavity field intensity $I(\bar{r})$. (b) Self-ordering of the density $n(\bar{r},t)$, numerically reconstructed from the particle trajectories. For blue detuned light ($\Delta >0$), atoms bunch in the minima of $I(\bar{r})$. The OAM dependent slope of the patterned phase in this graph corresponds to the rotation of the pattern in a ring geometry. Parameters are chosen as follows: OAM index $l$ = $3$, $\frac{\Gamma}{\kappa} = \bar{\omega}_p = \bar{\gamma} = 1,\, \zeta_p = 0.707,\, \Delta = 100$, such that $D'\approx 2,\, \sigma \approx 25$.}
\label{particle:fig}
\end{figure}

i.e. simply obtained as an ensemble average. Most importantly, this provides an expression of the relative constants in Eq. (\ref{smolu}) in terms of microscopic parameters:

\begin{equation}
D' = \frac{4 \omega_q^2}{\bar{\gamma}^2} D'_p = \frac{4 \omega_q^2}{\bar{\gamma}} \zeta_p^2, \quad \sigma = \frac{\omega_q\Gamma\Delta}{2\bar{\gamma}D'} = \frac{\Gamma \Delta}{8\kappa \omega_q^2 \zeta_p^2}
\end{equation} 

where the diffusion $D'$ and  $\sigma $ have been expressed in terms of the (scaled) momentum spread $\zeta_p^2$. Note that $D'_p = \bar{\gamma}\zeta_p^2$. 
\begin{figure}
\includegraphics[scale=0.27]{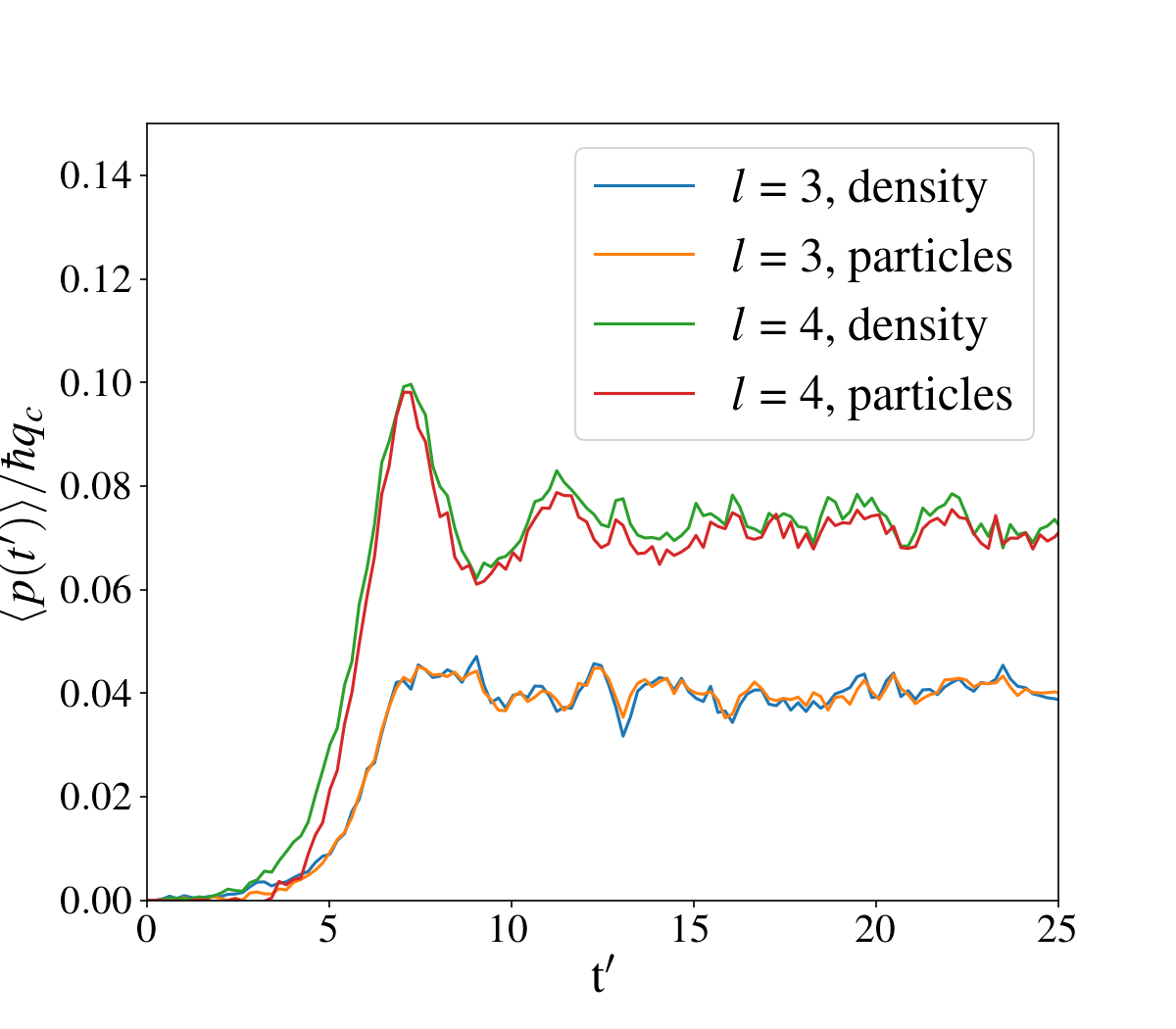}
\includegraphics[scale=0.42]{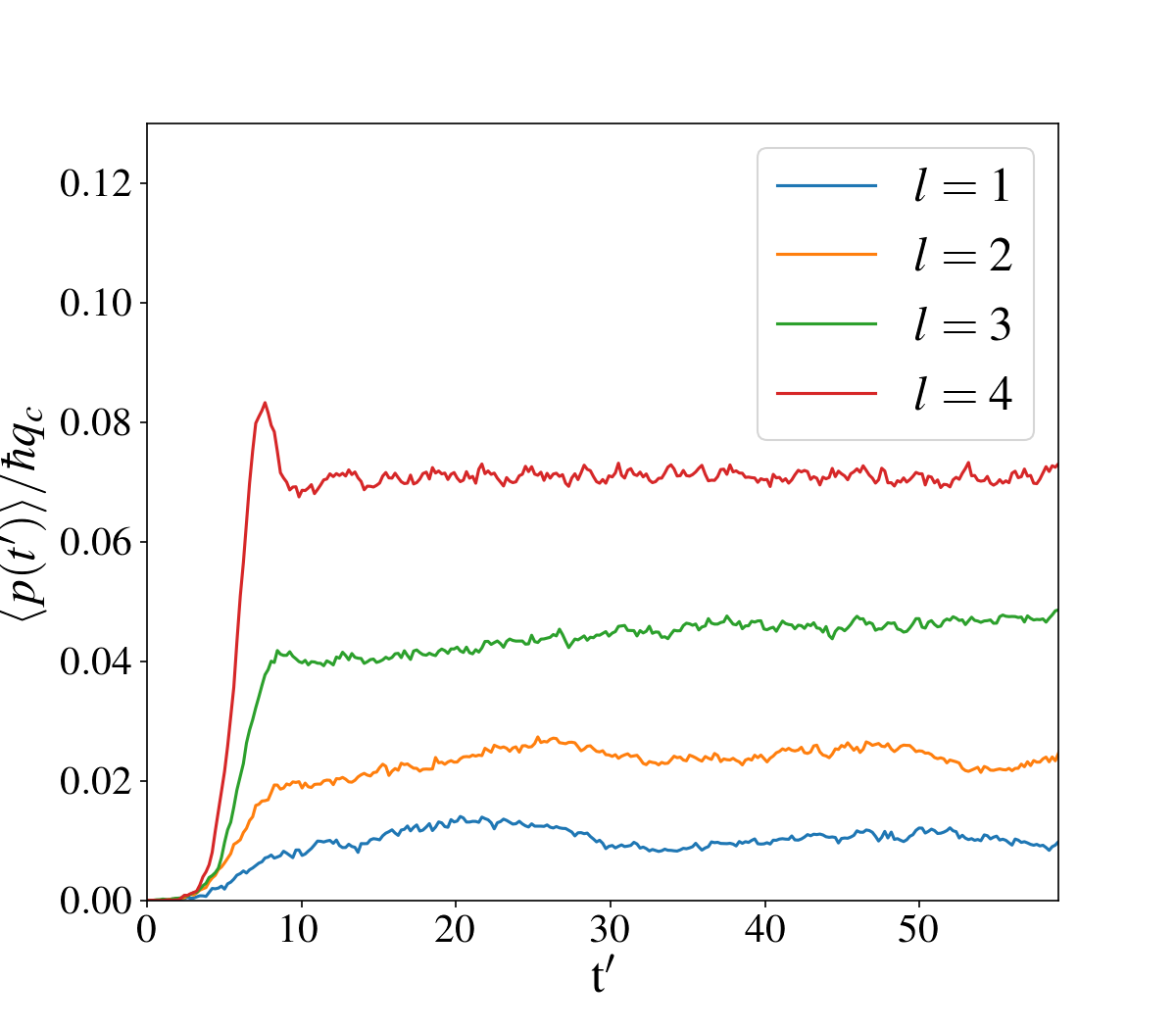}
\caption{\small Results from the 1D particle simulations. Atoms initially at rest and homogeneously distributed in the 1D domain display a non zero value of atomic current at the onset of self-organization. (Top) Time evolution of the atomic current $\langle \bar{p}_j(t')\rangle$, in units of $\hbar q_c$, with OAM index $l=3,4$, comparing the values arising from the ensemble average in Eq. (\ref{finmom}) and from the reconstructed density profile in Eq. (\ref{finmompart}). (Bottom) Evolution to steady state values of the atomic current, in units of $\hbar q_c$ at different values of the OAM index $l$. Each plotted line is in turn obtained by averaging over a set of 10 identical launches for each $l$. This results demonstrate the presence of atomic transport along the 1D azimuthal rings, induced by the pump OAM.}
\label{particle2:fig}
\end{figure}
We now integrate numerically the particle Eqs. (\ref{stoch1}) in the overdamped limit. Those are coupled to the field equation in 1D through a density profile $n(\bar{r},t')$, numerically reconstructed at all times from the particle positions $\bar{r}_j(t')$.

 A typical outcome of our simulations is presented in Fig. (\ref{particle:fig}), for the case of $5\times 10^4$ atoms. Consistently with the 2D numerical integration in Sec. \ref{sec:level3}, we observe drifting pattern dynamics in the 1D azimuthal geometry, i.e. ring lattice rotation at a fixed radius, induced by the OAM carried by the input pump. 
The results displayed in  Fig. (\ref{particle2:fig}) show that, at the onset of self-organization\footnote{This happens at roughly $t' = \kappa t \approx 5$ in our case.}, the atoms spontaneously develop net mass current along 1D rings, in the presence of OAM. This corresponds to the slope (drift) of the patterned field intensity $I(\bar{r},t')$ and atomic density $n(\bar{r},t')$, displayed in the $t'$ vs. $\bar{r}$ plots in Fig. (\ref{particle:fig}). 
In particular, in Fig. (\ref{particle2:fig} - Top), we compare the values of the current $\langle \bar{p}_{j}(t') \rangle$ (in units of $\hbar q_c$) obtained by means of both the above approaches, i.e. from Eq. (\ref{finmom}) (density) and Eq. (\ref{finmompart}) (particles). In both cases ($l=3,4$), one observes a exponential-like increase of $\langle \bar{p}_{j}(t') \rangle$, which eventually evolves to fluctuations around a steady state value. Moreover, the two series show a highly correlated behaviour.  Finally, as shown in Fig. (\ref{particle2:fig} - Bottom), in order to obtain more precise values of the steady state currents $\langle \bar{p}_{j}(t') \rangle$ for different OAM indices $l=1,\dots,4$ , we averaged over a set of 10 simulations with the same set of parameters for each value of $l$. As a final remark, we observe from Fig. (\ref{particle2:fig} - Bottom), that the steady state current values increases non linearly with the OAM index $l$. This might suggest that the optomechanical transfer of OAM considered here, i.e. from the cavity field to the atoms, occurs more efficiently with higher order OAM modes input. However, a detailed analysis of this mechanism in such a configuration is out of the aim of this work and will be investigated in future studies.

\section{Concluding Remarks}

Clouds of cold atoms in optical cavities, under the action of a coherent beam of light carrying OAM, can spontaneously form concentric rings of rotating light-atom lattices in the transverse plane, by means of optomechanical self-structuring (See Fig. (\ref{patterns})). The rotation of these spatio-temporal structures is due to the gradient of the transverse phase distribution externally imposed on the pump beam. The observed angular velocity is directly proportional to the OAM and inversely proportional to the square of the ring radius (See Fig. (\ref{agr})). Moreover, for increasing friction and decreasing diffusion, the atoms introduce a drag that slows down the rotation of the patterned rings (See Fig. (\ref{difcof})). By means of numerical evidences, we showed that such ring lattice rotation can sustain macroscopic atomic motion along azimuthal 1D rings in the 2D transverse domain (See Figs. (\ref{particle:fig}, \ref{particle2:fig})). Therefore,  optomechanical transport and azimuthal mass currents induced by the OAM in the structured pump have been predicted for a cold thermal gas in the assumption of overdamped motion. 

A possible generalization of the case considered here is that of rotating  optomechanical instabilities without momentum damping \cite{tesio2}.  Similarly, extensions to the dynamics of optomechanical dissipative solitons and their interaction in the presence of an OAM carrying pump are also of interest for future theoretical studies and experimental realizations. Finally, our results can be applied to the case of a quantum degenerate gas, in order to study vortex formation and turbulent regimes \cite{turb}. The OAM transfer mechanism discussed here represents a potential platform to induce quantized persistent currents of ultra-cold atoms, where such states are indeed an ideal candidate for the realization of atomtronic devices in circular 1D atomic traps \cite{alison1, sonja1}.

\begin{acknowledgments}

We are grateful to T. Ackemann and S. Franke-Arnold for helpful discussions. All authors acknowledge financial support from the European Training Network ColOpt, which is funded by the European Union (EU) Horizon 2020 programme under the Marie Skłodowska-Curie Action, grant agreement 721465. 

\end{acknowledgments}

\end{document}